# Detecting Domain-Generation Algorithm (DGA) Based Fully-Qualified Domain Names (FQDNs) with Shannon Entropy

Adam Dorian Wong

(*MalwareMorghulis*)

Dakota State University

INFA 723 Cryptography

Final Project

13 MAR 2021

Adam Wong
INFA 723-D01
Final Report

## Detecting DGA with Shannon Entropy

**Abstract**

The Solarwinds intrusion of 2020 is one of the most devastating and wide-spread attacks across the United States (US). An unknown threat actor leveraged a supply-chain attack against Solarwinds to push malicious updates to end-organizations and infect a significant proportion of companies across the world. Advanced Persistent Threats (APTs) leverage novel methods for beaconing and exfiltrating data from victim networks, especially over protocols such as DNS. DNS is an antiquated protocol, but is still reliable and used by both general users and malicious actors. One of the malware-strains related to the Solarwinds attack was dubbed SUNBURST by FireEye. SUNBURST relies upon domain-generation algortithm (DGA) to construct long-string URLs in which leverages DNS for its primary command-and-control (C2) beacon. This is a critical weakness in the SUNBURST malware and adversary's tactics, techniques, and procedures (TTPs) which can be detected by Shannon Entropy. As such, the Shannon Entropy Formula can be codified in Python and become an exportable script to provide to network defenders to augment their existing defenses without complex or expensive appliances.

**Introduction**

Security appliances, machine-learning, and concepts such as Security Orchestration, Automation, & Response (SOAR) are never meant to replace defense analysts completely. These are merely tools intended to augment defense and reduce time wasted on engaging trivial problems. With respect to the SUNBURST malware, adversaries require C2 beacons to maintain

13 MAR 2021

hooks within a compromised network. Threat actors need some mechanism to pivot and laterally move as well as maintain their malware on victim systems. In SUNBURST, threat actors leveraged primary beacons over DNS prior to moving to alternate protocols or communication channels and Stage-2 activities. However, prior to moving to more covert communication channels in Stage-2, the adversary leveraged unique string beacons to ***.avsvmcloud[.]com***. Companies like McAfee and phone apps leverage long string URLs for telemetry.

When considering web proxies, analyzing domains becomes a problem of scale. A defense analyst's time is precious because they are fighting to overcome the adversary's *breakout time*, in which CrowdStrike's Global Threat Report coins as the time-gap between first compromise to successful lateral movement [**1**]. Shannon Entropy is a calculation to quantify probability, randomness, and uncertainty in strings of text or of information [**2**]. Shannon Entropy can be used to reasonably reduce the pool of suspicious long-string URLs to examine.

The premise is to examine the "*average*" URL queried in a DNS log and compare URLs based on the randomness or uncertainty packed into each string. Domain names can be quantified as Shannon Entropies and further discriminated with single-variable *t-tests*. The end-state is to prove that simple Python scripts can be a viable path to constructing: an easily exportable, open-sourced, written in a common or popular scripting language, and homebrew detection technique without having to subscribe to high-cost appliances and SIEMs.

### Domain Name Service

***RFC***. According to the IETF RFC1035 (DNS), domain names must follow the same formatting as APRANET hostnames. Domain names can only consist of numbers, alphabetical characters, hyphens, and within 63 characters [**3**]. DNS is a best-effort distributed hierarchical protocol. It leverages UDP/53 to translate domain names into IP addresses and in-reverse. DNS can also use TCP/53 for zone updates. DNS as a protocol was designed with abstraction in-mind – that the end user should not be aware of it happening at all.

**DNSSEC**. DNSSEC is intended to provide stronger authentication and integrity through public-key cryptography or public-key infrastructure (PKI). It is intended to mitigate attacks like DNS cache-poisoning by providing validation that the DNS resolutions came from an authentic origin or resolutions were not tampered in-transit [**4**]. The capability gap in DNSSEC and HTTPS certificate chains is: *what if threat actors leverage generate authentic certificate chains or PKI*? Similar to reflecting upon command-and-control (C2) activity: *what if the data in-transit was malicious to start with - how would we even know*?

**Threat vector**. Infoblox has defined eight different types of attacks leveraging DNS as hostile infrastructure: DDoS, Floods, DNS Tunneling, NXDOMAIN, Phantom Domains, Random Sub-Domain (Slowdrip), Domain Lockup, and Botnet Bot [**5**]. Other lower-level attacks exist such as cache poisoning.

**Telemetry**. DNS has been used for alternate purposes than strictly resolution of IP addresses. DNS is often used in telemetry. It can bypass firewalls subverting network-level defenses because traffic is destined for Port 53 and sensitive data can be encoded in prefixes or subdomains. Depending on the perspective, it's known as DNS Tunneling [**6**]. An example would be: McAfee Global Threat Intelligence which supplements the McAfee suite and crowd-sources threat data from clients [**7**] [**8**].

| Example Vendor Domains |
| --- |
| 4z9p5tjmcbnblehp4557z1d136.avqs.mcafee[.]com |
| sfqpit75pjh525siewar2dtgt5.avts.mcafee[.]com |

*Figure 1: Example telemetry from the McAfee Global Threat Intelligence service (source: McAfee).*

**Why DNS**? DNS is dangerous because it's an antiquated protocol and transparent to the user. All networks rely upon DNS to function. Therefore, it cannot be reasonably blocked at a network-level firewall. Risk can be managed through using specific DNS servers. Privacy and security-centric services like Quad9 (IP: 9.9.9.9) augments name resolutions using threat intelligence-based DNS blocklists [**9**]. However, the capability gap remains between when a malicious website is reported to when a blocked URL is actually updated or implemented.

**Case Study: ExploderBot**. An open-source report produced by the US National Security Agency (NSA) highlighted an interesting observed attack over DNS. The attack was identified as a global threat dubbed: *ExploderBot*, which leverages a DDoS by way of slow-drip water torture attack against open-resolvers. With ExploderBot, the URLs used in the attack appear to have a diverse range of long or short fully-qualified domain names (FQDN) with varying lengths of subdomain prefixes and many of which reportedly have origins in China [**10**]. An interesting item to note is these subdomains are part of intricate calculations within multiple layers: DNS, IP, and UDP headers [**10**]. DNS is dangerous when misused. For this case study, ExploderBot attack leverages DNS, it was near impossible to detect, and presumably difficult to develop detection signatures because activity was buried deep within the bytes of headers. Attacks are becoming more sophisticated, quieter, and harder to detect. But at the core of it, adversaries need to blend-in and generate network traffic consistent with normal behavior. To do so, they rely on DNS to communicate beyond the compromised network, therefore they can be detected.

## Cyber Kill Chain

***Kill Chain***. A time-honored measure of adversary progression is the Lockheed Martin Cyber Kill Chain© (Kill Chain). The Kill Chain is divided into linear phases that represent the intrusion from reconnaissance, weaponization, delivery, exploitation, installation [of persistence], command-and-control (C2), and actions-on-objective [**11**]. Now, a Lockheed Martin team propose that network defenders have multiple courses-of-action (COAs): detect, deny, disrupt, degrade, deceive, and destroy [**11**].

***C2***. Kill Chain (Phase Six, KC6) C2 gives us the last opportunity to observe activity, apply mitigations, and to push-back against an adversary's complete dominance in the compromised network. Malware needs a means to control malware, maintain persistence, laterally move, and bridge adversary to the victim infrastructure [**12**]. SANS Instructor, Dr. Eric Cole states: "*Prevention is ideal but detection is a must - hunting, detecting and controlling the damage of an attack is critical in reducing risk...*" [**13**]. That mission requirement for external communication is a weakness for threat actors.

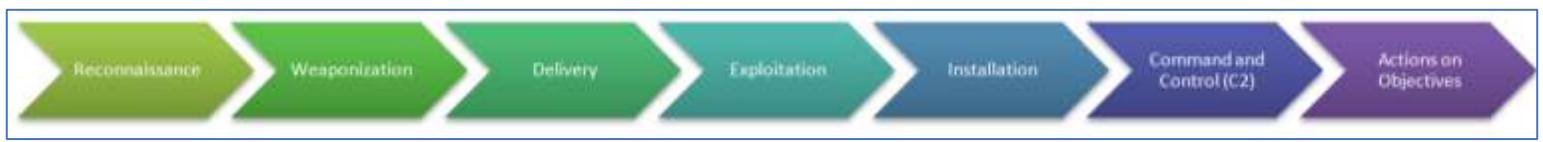

*Figure 2: Cyber Kill Chain (image source: countuponsecurity.com)*

## SUNBURST Intrusion

In December 2020, FireEye responded to an internal intrusion. During FireEye's internal investigation, analysts found that its proprietary Red Team tools were stolen by an undisclosed threat actor. Once FireEye ascertained the scope of the intrusion, the rest of the country found itself also affected by the compromise of Solarwinds appliances. Now, for the purposes of the analysis, the Diamond Model of Intrusion Analysis will be used to analyze the attack [**14**].

*Adversary*. FireEye dubs the threat actor group as UNC2452 (Uncategorized) prior to be labeled or attributed to any Advanced Persistent Threat (APT) or Financial Crime (FIN) group. The adversary was able to infect several versions and hotfixes of the Solarwinds Orion Platform. Although it remains as speculation, JetBrains TeamCity software has been suspected of providing an avenue to merge the malware into the Solarwinds production code [**15**]. CISA asserts that the campaign began as early as March 2020 [**16**]. FireEye noted the adversary relies on legitimate credentials to pivot within the victim network [**17**]. Therefore, the actor is harder to detect once they're inside the compromised victim network – the ingress-egress traffic is paramount to identifying them.

*Capability*. The initial entry vector (IEV) for the SUNBURST malware was through a malicious dynamically-linked library (DLL) called: *solarwinds.orion.core.businesslayer.dll*. When initiating C2 callbacks, the malware attempts to masquerade as Solarwinds telemetry behavior, more specifically the Orion Improvement Program (OIP) protocol [**16**] [**17**]. The malware itself leverages domain-generation algorithms (DGA) to encode information about the victim network. The data is encoded into DNS resolutions of victim-specific DGA-based URLs to: ***.avsvmcloud[.]com**. CISA identifies affected organizations as either: Category-2 (malicious DLL exists, primary C2 beaconing may exist) or Category-3 (malicious DLL exists, and beaconing swapped from primary to secondary URLs or IPs). The prefixes when unscrambled include sleep timers to notify a compromised server how long to pause DNS beacons [**18**]. Data, system information, hostnames, even anti-virus information was hidden inside the subdomain prefixes leveraging a combination of encryption of FNV-1a and XOR functions [**17**].

| Example SUNBURST Primary C2 Domains |
|---|
| <SUBDOMAIN>.appsync-api.eu-west-1.avsvmcloud[.]com |
| <SUBDOMAIN>.appsync-api.us-west-2.avsvmcloud[.]com |
| <SUBDOMAIN>.appsync-api.us-east-1.avsvmcloud[.]com |
| <SUBDOMAIN>.appsync-api.us-east-2.avsvmcloud[.]com |

*Figure 3: Sanitized C2 domains used as primary beacons for victim organizations affected by SUNBURST (source: FireEye).*

***Infrastructure***. The hostile primary C2 server (*avsvmcloud[.]com*) returns the DNS response of an IP address, in which case the victim switches to another hostile-server over encrypted communications to HTTPS. CISA notes that the last-mile or last-hop to the target network came from a diverse array of hop-points across the world and within the US [**16**].

***Victimology***. High-value targets were further infected by Stage-2 malware and directed towards secondary domains and implants [**16**] [**19**]. This demonstrates the tenacity of the threat group. The selective nature of deploying secondary beacons suggests the adversary's motive to prepare targets, provide effects, or collect intelligence on specific and noteworthy targets.

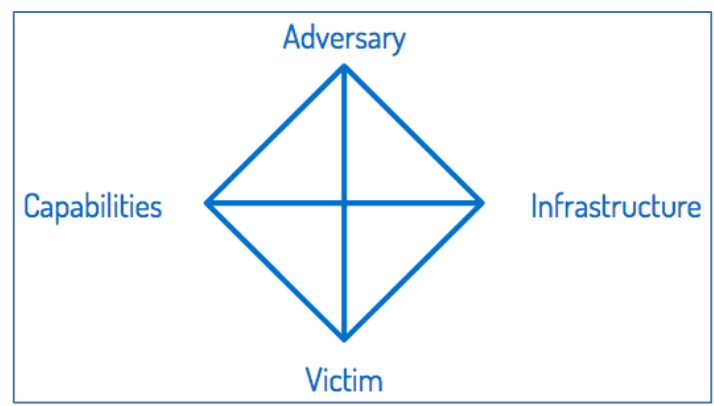

*Figure 4: Diamond Model of Intrusion Analysis (image source: RecordedFuture).*

Therefore, actors responsible for the Solarwinds intrusion are highly dangerous, patient, and technically proficient. However, they can be detected based on their use of DGA or long-string domain names in their primary C2 DNS beacons.

**Indicators of Compromise**

The *Pyramid of Pain* was designed by David Bianco. The premise of the logical framework is that Indicators-of-Compromise (IOC) have different weights and worth to an adversary. In order to induce the most pain to an adversary's offensive operation, defenders must nullify higher worth indicators such as toolkits or tactics, techniques, and procedures (TTPs) [**20**]. This threat actor used DGA for their first efforts in reach-back. This is not just a domain name indicator that easily expires. To the contrary, this is a TTP to pivot on because it is commonality within this specific intrusion across multiple victims.

**Case for Using Shannon Entropy**

The intent is to break the cyber kill chain by applying detective techniques in the earliest possible phases of the intrusion. For SUNBURST, the intrusion was nearly impossible to detect because the malware was injected via supply-chain attacks and leveraging known-good signing certificates. Therefore, the best chance of detection is at the network-level because it can scale to the problem and affected victim infrastructure. The premise is that Shannon Entropy helps quantify the uncertainty and chaos in information, the higher the entropy, the less likelihood that it is normal to defenders [**21**].

$$H(X) = H(p_1, ..., p_n) = -\sum_{i=1}^{n} p_i \log_2 p_i$$

*Figure 5: Shannon Entropy Formula (image source: TowardsDataScience.com)*

***DNS Tunneling***. Use of Shannon Entropy to detect DNS Tunneling is not a new concept. Security researchers have used Shannon Entropy to identify malicious dynamic DNS hostnames [**22**]. A Splunk extension called: "*URL_Toolbox*" which provides the capability to feed a domain name from a search query into a function and generates a Shannon Entropy field [**23**].

**Proprietary Tools**. As previously stated, appliances like Splunk have extensions to provide similar functionality. However, cost is a significant issue, especially for smaller organizations. Therefore, a lower-cost solution is necessary to provide enough diffusion and detection capabilities to a wider audience. The solution needs to be simple, open-sourced, and effective.

## Limitations Using Shannon Entropy

In any solution that is designed, it is imperative to consider the data itself as a constraint. Data changes and the support for it changes. These may affect the accuracy and precision of Shannon Entropy. It becomes a risk-appetite issue for the end-user of the entropy detection tool.

**Geographic Considerations**. A strategic-level problem comes from during data set selection. *Which Top-1M websites are being considered*? The Top-1 Million websites are different between different countries, languages, which may affect character frequency tables.

**Data Set Size**. The natural operational limitation of Shannon Entropy derives from the sample used in building character probability tables. Alexa Top-1M is now owned by Amazon; it shows the Top-500 websites with all additional data sets requiring subscription [**24**].

**Data Structure**. The file formats which domains are ingested into any script will need to be routinely monitored because it can break scripts or distort the Shannon Entropy calculations.

**Source Lifecycle**. Some vendors offer Top-N website data for free such as with Cisco Umbrella and Majestic Million [**25**] [**26**]. It's unclear when a vendor will abruptly cease support of providing the data.

*Character Probability*. A tactical-level problem in development was capturing an appropriate data set to build a character probability table. It also relied on the issue of scope: *what do we consider probability of a letter occurring? The next letter after a character based on the English dictionary? Or character frequency?* To reduce the scope, probabilities generated by the project tool was to use a generic sample with character probabilities which are based frequency within the sample data set. It's an imperfect method, because it does not take languages into account. However, RedCanary has used similar techniques to build their character probability table [**27**].

## Demonstration of the Tool

**Log Source**. For testing purposes, DNS log data was extracted from a Pi-Hole DNS server. Some data was sanitized for home network operational security reasons, which may affect the entropy value itself. However, the input data is a simple text file that strictly contains the fully-qualified domain names (FQDNs), one per-line.

```
FQDN is: xp.apple.com; Length is: 13; Entropy is: 0.6086892572844841
FQDN is: xsts.auth.xboxlive.com; Length is: 23; Entropy is: 1.1088998509145904
FQDN is: ykpxeewrtzypdrg.lan; Length is: 20; Entropy is: 0.9575760732565867
FQDN is: zn42v6draxyafsjmv-homedepot.siteintercept.qualtrics.com; Length is: 55;
 Entropy is: 2.9783271930544615

Mean URL Length is: 31.244898, StdDev is: 16.802196
Mean Entropy is: 1.263970, StdDev is: 0.504894
>>>
```

*Figure 6: Phase 1 - shannon.py (v1.0), proof-of-concept which uses hardcoded filenames and values from RedCanary's static character table. It simply returned entropy values for all domains. Phase 1 – minor tuning to narrow suspicious domains (v1.1).*

*Phases 1 & 2*. The tool was developed in several phases (Figure 6). The first iteration (Phase 1 in v1.0 and Phase 2 in v1.1) were simple scripts that leveraged the RedCanary's static character frequency table [**27**]. Reiterating the problem that the data is stale, Top-N websites change. However, this provided a proof-of-concept that using Shannon Entropy can work in Python. With the Phase 0 script, it simply ran by itself with a hardcoded file name and values, expecting an input file called "*domains.txt*" to simply exist in the same folder. Output files were generated as well as files to Standard-Output (STDOUT). Testing between Phases 0 and 1 were often blurred due to iterative testing for each module or function added to the script. During these phases an arbitrary value was used to mark Shannon Entropy values.

  *Phase 3*. The script *dictionary.py* was created to conduct on-the-fly testing of modules to be added into the production script. The intent was to test capability of using more Top-N sources other than a static table. Those modules were later merged into the production script. For example, using Cisco's Top-1M or Majestic Million which are constantly refreshed [**25**] [**26**].

  *Phase 4*. The production script (v2.0) Shannon Entropy values were provided and an arbitrary test was established to seek values greater than two standard deviations above the mean (instead of an arbitrary value for development and testing purposes). The production script

```
Shannon_Entropy: 3.5898735065950667; Suspect_FQDN: clickstream-producer.hd-personalization-prod.gcp.homedepot.com
Shannon_Entropy: 3.6591188110369175; Suspect_FQDN: 23o27zncnk4smebaxlzwpnoc23vzt2bpbjrknvf7b613f18efab976d6sac.d.aa.online-metrix.net
Shannon_Entropy: 3.6985100501189954; Suspect_FQDN: 23o27zncgqftu6t6sppduslkbrpejbw5cpunimac67580b1f644d017dsac.d.aa.online-metrix.net
Shannon_Entropy: 3.7032700268766354; Suspect_FQDN: 23o27zncs4b6gsfs4wlmeah5yzniq5zgrqawmbpo0f29cffdfbd747d1sac.d.aa.online-metrix.net

### Entropy Test ###
Shannon_Entropy > 2.2147400154696468 is statistically significant!

### Averages ###
Avg URL Length is: 31.244898, StdDev is: 16.802196
Avg Entropy is: 1.232662, StdDev is: 0.491039
Avg Prefix Length is: 17.531250, StdDev is: 11.927156
```

provides limited functionality for a user to set specific files instead of hard-coded values. It also gives the capability to select the desired Top-N file (Cisco, Majestic, or default: RedCanary).

*Figure 7: Production version of the shannon.py solution which relies on the data soruce to provide the "cut-off" mark or waterline for suspicious high-entropy domains.*

***Strength***. The testing of the script was successful in discriminating between FQDNs which passed the Shannon Entropy test (falling within two standard deviations) and those that failed in a Windows environment in both Python IDLE and Visual Studio Code terminals (Figure 7). It detected suspicious DGA activity and provided back a reasonably-sized comma-separated value (CSV) file of suspicious domains to re-examine compared to the full list. Two standard deviations are adequate for hunting purposes because according to a generic normal model in statistics. Two standard deviations should represent the middle-95% of the entire data set. Testing occurred under the simulation that a SUNBURST primary C2 domain was logged and being scrutinized.

*Figure 8: The script successfully detected the SUNBURST DGA-based primary C2 beacon URL in RED, with the standard for detection in YELLOW (URL source: securelist.com).*

***Weakness***. A weakness in the script includes lacking a mechanism to whitelist or permit previously observed high-entropy domains. It became a problem of scale and possibly solving regular expressions. This script executes a simple single-variable *t-test*, it can be improved by adding confidence intervals or a means to modify the upper-bound mark and provide the end-user the agency to select the waterline instead of the default two standard deviations. A counterargument to this self-critique is that defense analysts may not care enough about statistics and merely agree that at least two standard deviations difference is a good-enough waterline. In cases like ExploderBot, this script may have difficulty detecting smaller FQDNs observed in that attack. The script also lacked adequate change control which slowed, but did not inhibit development.

**Conclusion**

Given a large enough DNS log, the script is capable of detecting SUNBURST domains based on a combination of Shannon Entropy and a simple single-variable statistical *t-test*. SUNBURST relied on domain generation algorithms (DGA) to maintain its primary C2 activity with an infected host. The script is lightweight, utilizes several import modules in Python 3, and is interoperable across at least two frequently updating and major Top-N vendors. The script relies on few built-in dependencies, making it agile to move between systems with an existing installation of Python. However, through testing, the script has setbacks such as a lack of whitelisting capabilities and cannot detect attacks using smaller URLs as seen in ExploderBot. Shannon Entropy is an imperfect solution to a complex problem. Data sources can change which adds a layer of complexity to the script. Character frequency charts were derived from Top-N websites. These also yield complexity because data sources, data formatting, and data geographical relevance naturally affect the tool. Overall, the use of Shannon Entropy can help detect low-profile C2 channels and SUNBURST domains based on statistical analysis of the uniquely-long FQDNs.